\documentclass[aps,twocolumn,showpacs,showkeys,amsmath,amssymb,superscriptaddress,nofootinbib,floatfix,longbibliography]{revtex4-2}
\usepackage{amsmath,graphicx,color,ulem,float}
\usepackage{epsfig,bm,slashed,hyperref}
\newcommand{\trento}{T$\mathrel{\protect\raisebox{-2.1pt}{R}}$ENTo}

\begin{document}

\title{Enhanced hydrodynamic predictions for $v_{02}(p_T)$}

\author{Rupam Samanta}
\affiliation{Institute of Nuclear Physics, Polish Academy of Sciences, 31-342 Cracow, Poland}
\author{Tribhuban Parida}
\affiliation{AGH University of Krakow, Faculty of Physics and Applied Computer Science, aleja Mickiewicza 30, 30-059 Cracow, Poland}
\author{Jean-Yves Ollitrault}
\affiliation{Universit\'e Paris Saclay, CNRS, CEA, Institut de physique th\'eorique, 91191 Gif-sur-Yvette, France}

\begin{abstract}
  We present hydrodynamic predictions for the new observable $v_{02}(p_T)$, which measures the correlation of particle spectra with elliptic flow. 
  We implement a data-driven correction so as to match hydrodynamic calculations to elliptic flow ($v_2(p_T)$) data. 
  The corrected results are in fair agreement with $v_0(p_T)$ data up to high $p_T$. 
  We make predictions for $v_{02}(p_T)$ of unidentified charged hadrons up to $p_T=10$~GeV$/c$, and of pions, kaons and protons up to $p_T=5-6$~GeV$/c$, in several centrality windows, for Pb+Pb collisions at $\sqrt{s_{NN}}=5.02$~TeV. 
  We predict a decrease of $v_{02}(p_T)$ of charged hadrons for $p_T>4$~GeV$/c$ and meson-baryon splitting for $p_T>3$~GeV$/c$ in mid-central collisions. 
  We also predict a non-monotonic variation of $v_{02}(p_T)$ for protons at low $p_T$ above $30\%$ centrality. 
  This is a specific feature of this new observable, which is not observed for the usual flow observables $v_2(p_T)$ and $v_0(p_T)$. 
 \end{abstract}

\maketitle

\section{Introduction}
\label{s:intro}

The collective nature of the radial expansion of the quark-gluon plasma (QGP) produced in ultrarelativistic heavy-ion collisions is probed by the observable $v_0(p_T)$~\cite{Schenke:2020uqq}, which has recently been measured in Pb+Pb collisions at the Large Hadron Collider (LHC)~\cite{ATLAS:2025ztg,ALICE:2025iud}.
$v_0(p_T)$ is (up to a normalization factor), the relative change in the transverse momentum ($p_T$) distribution of outgoing particles (hereafter referred to as ``spectrum'') generated by a small increase in the QGP temperature~\cite{Parida:2024ckk}. 
We have recently proposed a similar new observable $v_{02}(p_T)$~\cite{Parida:2025eqv}, which is the relative change in the spectrum generated by a small increase in elliptic flow ($v_2$). 
Elliptic flow measures the geometric shape of the QGP, and its effect on the spectrum has already been revealed using event-shaped-engineered events~\cite{ALICE:2015lib}, thus providing a first indirect measurement of $v_{02}(p_T)$~\cite{Parida:2025eqv}. 
We anticipate that dedicated analyses of $v_{02}(p_T)$ will soon be carried out using RHIC and LHC data.
The advantage of this new observable over $v_{0}(p_T)$ is that it is measured through a three-particle cumulant, as opposed to a two-particle correlation. 
Therefore, it is less sensitive to nonflow effects, which are a well-known bias in peripheral collisions or at high $p_T$.
On the other hand, $v_{02}(p_T)$ is not trivially related to $v_0(p_T)$ because the change in the spectrum measured by $v_{02}(p_T)$ contains a purely geometric component~\cite{Samanta:2026omo}, in addition to the thermal component probed by $v_0(p_T)$. 

In this paper, we make predictions for $v_{02}(p_T)$ for Pb+Pb collisions at LHC energies, with the aim of being  realistic enough that they can be readily compared with the first experimental data when available. 
We expect that  $v_{02}(p_T)$ will be measured for charged hadrons and identified particles for $p_T$  up to a few GeV$/c$  in several centrality windows, in the same way as $v_0(p_T)$~\cite{ATLAS:2025ztg,ALICE:2025iud}, and our predictions span the same range. 
We use hydrodynamics, which is the natural framework for modeling elliptic and radial flow~\cite{Kolb:2003dz}. 
We model event-by-event fluctuations in the initial-state \cite{Hama:2004rr,Holopainen:2010gz}, which are essential for the observable under study. 
But at variance with usual state-of-the-art models~\cite{Bernhard:2019bmu,Nijs:2020ors,JETSCAPE:2020shq}, we run {\it ideal\/} hydrodynamics rather than {\it viscous\/} hydrodynamics. 
The motivations for this lack of sophistication are detailed in Sec.~\ref{s:correction}. 
Ideal hydrodynamics overestimates elliptic flow~\cite{STAR:2000ekf,Romatschke:2007mq}, in a way which depends non-trivially on centrality, $p_T$ and particle species, as will also be shown in Sec.~\ref{s:correction}. 
We model this effect crudely as a suppression factor $f(p_T)$ which we assume to be identical for $v_2(p_T)$,  $v_0(p_T)$ and $v_{02}(p_T)$, in the spirit of a core-corona separation~\cite{Hirano:2005wx,Werner:2007bf,Kanakubo:2021qcw} in which only a fraction $f(p_T)$ of the particles belong to the fluid. 
We evaluate $f(p_T)$ using $v_2(p_T)$ data.
In Sec.~\ref{s:predictions}, we show that  $v_0(p_T)$ data are well understood on this basis, and we predict $v_{02}(p_T)$ along the same lines.

\section{Hydrodynamic simulations and data-driven correction}
\label{s:correction}

We simulate Pb+Pb collisions at $\sqrt{s_{NN}}=5.02$~TeV, which is the energy at which $v_0(p_T)$ has been analyzed~\cite{ATLAS:2025ztg,ALICE:2025iud}. 
Note that collisions are now carried out at a slightly higher energy, $5.36$~TeV~\cite{CMS:2024ykx}, but this  is likely to have a negligible effect on the observable studied here. The details of our simulation framework are briefly summarized in Appendix~\ref{s:hydrodetails}. Notably, the medium evolution is performed using ideal hydrodynamics.

We explain our motivation for running ideal hydrodynamics, rather than viscous hydrodynamics, which is the standard approach for heavy-ion phenomenology~\cite{Bernhard:2019bmu,Nijs:2020ors,JETSCAPE:2020shq}. 
Ideal hydrodynamics has fewer parameters and is better constrained theoretically: In particular, hadrons are emitted at freeze-out according to boosted thermal distributions~\cite{Cooper:1974mv,Andronic:2017pug}, so that there is no ambiguity on how observables are calculated. By contrast, in viscous hydrodynamics, hadronization is poorly understood: 
How the off-equilibrium part of the momentum distribution is distributed as a function of momentum~\cite{Teaney:2003kp} depends on the details of the microscopic interactions~\cite{Dusling:2011fd,Molnar:2014fva,Plumari:2015sia} which results in an ambiguity when evaluating quantities such as $v_2(p_T)$~\cite{JETSCAPE:2020mzn}. 
 
\begin{figure}[!htbp]
    \centering
\vspace{-25pt}
\includegraphics[width=\linewidth]{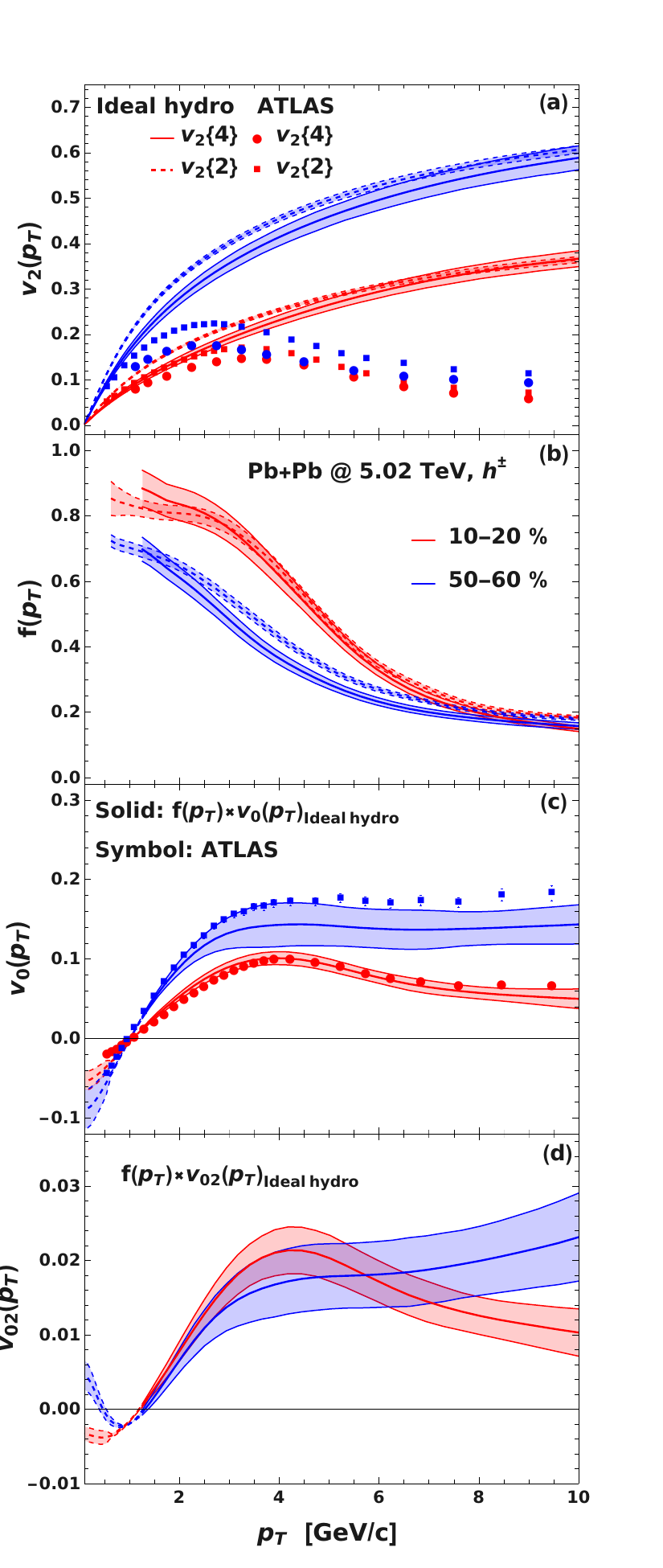}
    \caption{
    (a) Symbols: ATLAS data for $v_2\{2\}(p_T)$ and $v_2\{4\}(p_T)$ of charged hadrons~\cite{ATLAS:2024mch} in two centrality windows. 
    Lines: our ideal hydrodynamic calculation. 
    (b) Correction factor $f(p_T)$, defined by Eq.~(\ref{defcorrection}), evaluated using $v_2\{2\}$ (dashed lines) and $v_2\{4\}$ (full lines). 
    (c)  Symbols: ATLAS data for $v_0(p_T)$ of charged hadrons~\cite{ATLAS:2025ztg}, with $0.5<p_T^{ref}<2$~GeV$/c$ (see text). 
    Lines:  our hydrodynamic calculation multiplied by $f(p_T)$. 
    (d) Our prediction for $v_{02}(p_T)$ of charged hadrons. 
    Bands  in panels (c) and (d) depict statistical error bars from the finite number of initial conditions, evaluated with the bootstrap method, and from the determination of $f(p_T)$. 
    }
\label{fig:ATLAS}    
\end{figure}

\begin{figure*}[!htbp]
    \includegraphics[width=\linewidth]{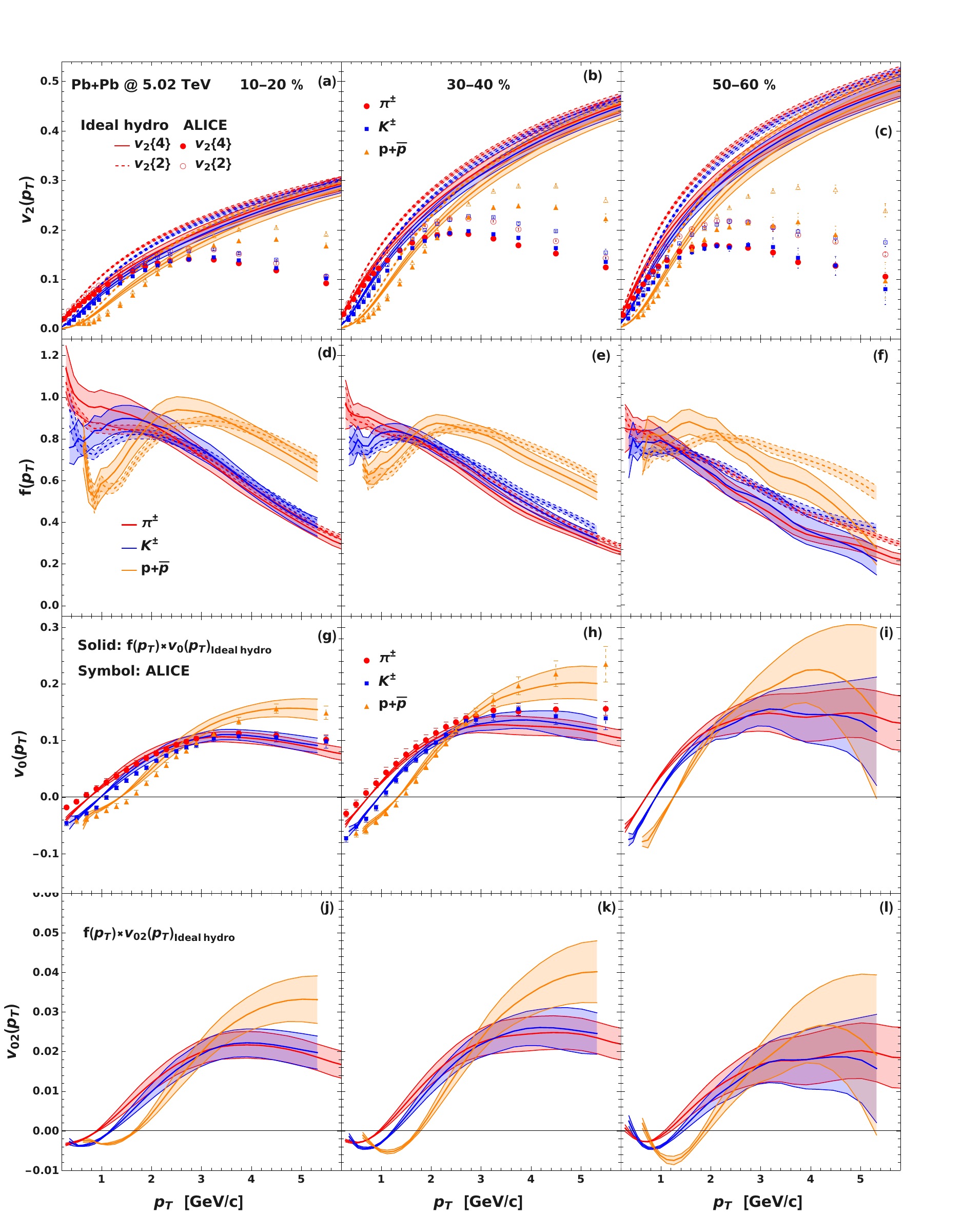}
     \caption{
    Same as Fig.~\ref{fig:ATLAS} for identified pions, kaons and protons, and for three centrality windows (left: 10-20\%, middle: 30-40\%, right: 50-60\%). 
    Symbols are ALICE data from Fig.3 of Ref.~\cite{ALICE:2022zks} ((a)-(c)) and from Ref.~\cite{ALICE:2025iud} ((g)-(h)). }
\label{fig:ALICE}
\end{figure*}

Despite the larger freedom (associated with transport coefficients, and the momentum distribution at freeze-out), viscous hydrodynamics alone is unable to reproduce a number of universal features seen at high $p_T$ (by which we mean  $p_T\gtrsim 3$~GeV$/c$), such as the decrease of $v_2(p_T)$ of charged hadrons~\cite{CMS:2012xss,ATLAS:2012at,ATLAS:2018ezv} (Fig.~\ref{fig:ATLAS} (a)), and the fact that $v_2(p_T)$ is larger for baryons than for mesons (meson-baryon splitting, Fig.~\ref{fig:ALICE} (a)-(c))~\cite{STAR:2003wqp,PHENIX:2003qra,ALICE:2013snk,ALICE:2014wao}. 
Interestingly,  $v_0(p_T)$ displays the same features~\cite{ATLAS:2025ztg,ALICE:2025iud} (Fig.~\ref{fig:ATLAS} (c) and  Fig.~\ref{fig:ALICE} (g)-(h)). 
They are generic properties of collective flow~\cite{ALICE:2021ibz,ALICE:2022zks}, and we expect that they will be also visible in $v_{02}(p_T)$. 
If we want to capture these phenomena, viscosity will be of little help. 
Therefore, we choose to run ideal hydrodynamics, for the sake of simplicity, and we then correct our results in a data-driven approach so as to capture the desired phenomena, as we now explain. 

Fig.~\ref{fig:ATLAS} (a) and Fig.~\ref{fig:ALICE} (a)-(c) display the results of our hydrodynamic simulation for $v_2(p_T)$, together with experimental data. 
There are several measures of $v_2$, depending on how many particles are correlated~\cite{Borghini:2001vi}. 
We display the estimates from 2 and 4 particle cumulants,  denoted by $v_2\{2\}$ and $v_2\{4\}$. 
We recall in Appendix~\ref{s:hydrodetails} how we evaluate these two observables in hydrodynamic simulations~\cite{Niemi:2015qia}. 
These simulations always predict $v_2\{2\}>v_2\{4\}$, where the difference is due to event-by-event $v_2$ fluctuations, and the same ordering is observed experimentally. 
In addition, the experimental $v_2\{2\}$ usually has a significant positive bias from nonflow correlations (typically back-to-back jets), which causes a deviation from hydrodynamic calculations where nonflow effects are not included. 

We start by discussing results for unidentified charged hadrons, shown in Fig.~\ref{fig:ATLAS} (a). 
In hydrodynamics, $v_2(p_T)$ first increases linearly~\cite{Borghini:2005kd} and then more slowly, displaying a negative curvature, which is a well-known effect of initial-state fluctuations~\cite{Andrade:2008xh}: 
Particle emission for $p_T\gtrsim 2$~GeV$/c$ typically comes from local ``hot spots''~\cite{Gale:2012rq}, whose shape is not correlated with the overall geometry, so that they do not contribute to $v_2(p_T)$. 
These are robust features of the hydrodynamic calculation, which do not depend on model details. 
Experimental data are systematically below hydrodynamic calculations. 
At low $p_T$, this is a well-known effect of viscosity, which suppresses $v_2(p_T)$ by a global factor, the suppression being stronger for more peripheral collisions~\cite{Gardim:2020mmy,Gardim:2022vys}. 
Above $p_T\approx 3$~GeV$/c$, $v_2(p_T)$ decreases with $p_T$. This decrease is observed for both $v_2\{2\}$ and $v_2\{4\}$, so that it cannot be ascribed to nonflow effects. 
It is not seen in hydrodynamic simulations and represents a generic discrepancy between data and hydrodynamics, which cannot be resolved by adjusting parameters.  

Let us now turn to results for identified hadrons, shown as lines in~Fig.~\ref{fig:ALICE} (a)-(c). 
At low $p_T$, hydrodynamics reproduces the mass ordering seen in data, that $v_2(p_T)$ at fixed $p_T$ decreases as the particle mass increases~\cite{Huovinen:2001cy}. 
This is also a robust feature of hydrodynamic modeling~\cite{Borghini:2005kd}, which is present both for $v_2\{2\}$ and $v_2\{4\}$. 
At higher $p_T$, however, this mass ordering is broken, and $v_2(p_T)$ becomes larger for protons than for pions and kaons. 
More generally, $v_2(p_T)$ is larger for baryons than for mesons~\cite{ALICE:2014wao}.
This trend, which is again seen for both $v_2\{2\}$ and $v_2\{4\}$, cannot be understood within hydrodynamics, where mass ordering is present for all $p_T$. 
The meson-baryon splitting is usually explained by invoking coalescence of constituent quarks into hadrons~\cite{Fries:2003vb,Molnar:2003ff,Fries:2003kq,Greco:2003mm}, which goes beyond the mere hydrodynamic description. 

We consider that the decrease of flow at high $p_T$ and meson-baryon splitting are non-hydrodynamic features. 
We model these effects heuristically by assuming that only a fraction $f(p_T)$ of the produced particles is described by hydrodynamics, while the remaining fraction $1-f(p_T)$ does not flow. 
We evaluate the fraction $f(p_T)$ using elliptic flow: 
\begin{equation}
\label{defcorrection}
f(p_T)\equiv \frac{v_2(p_T)[{\rm data}]}{v_2(p_T)[{\rm hydro}]},
\end{equation} 
where the numerator is from data, and the denominator from our hydrodynamic calculation. 
In other words, the correction factor $f(p_T)$ is defined as that which would restore perfect agreement between our hydrodynamic calculation and elliptic flow data. Since the denominator is obtained from ideal hydrodynamic simulations, $f(p_T)$ can be interpreted as an effective correction factor that accounts for both viscous effects and other non-hydrodynamic contributions at a given $p_T$.
We can use Eq.~(\ref{defcorrection}) for $v_2\{2\}(p_T)$ or for  $v_2\{4\}(p_T)$. 
We thus obtain two independent estimates of $f(p_T)$, and  include the difference between these estimates in the error bar of our predictions, as will be detailed below. 

%Using the ideal hydrodynamic calculation as the baseline, we extract the correction factor $f(p_T)$ by taking the ratio of the charged-hadron $v_2\{4\}(p_T)$ measured by ATLAS~\cite{ATLAS:2024mch} to the corresponding ideal hydrodynamic results. 
The extracted $f(p_T)$ is shown in Fig.~\ref{fig:ATLAS}(b) for charged hadrons in two centrality windows.
It is obtained by taking the ratio of the symbols (data) and lines (ideal hydrodynamics) in Fig.~\ref{fig:ATLAS}(a).  
$f(p_T)$ is always smaller than unity,  as expected since ideal hydrodynamics overestimates elliptic flow. 
Values of $f(p_T)$ obtained using $v_2\{2\}$ and $v_2\{4\}$ are in fairly good agreement. 
There is a sizable difference at high $p_T$, where the estimate from 2-particle cumulants is larger than from 4-particle cumulants. 
The difference is larger for more peripheral collisions. 
Both trends are compatible with the expectation from nonflow effects, which increase the numerator of Eq.~(\ref{defcorrection}) for $v_2\{2\}$. 
At low $p_T$, $f(p_T)$ is fairly close to unity for more central collisions, and smaller for more peripheral collisions. 
These features are consistent with the expectations from viscous suppression. 
If we had run viscous hydrodynamics with realistic transport coefficients, $f(p_T)$ at low $p_T$ would be close to $1$ for all centralities. 
Note that ATLAS data are only available for $p_T>1$~GeV$/c$ in the case of $v_2\{4\}$, and $p_T>0.5$~GeV$/c$ in the case of $v_2\{2\}$. 
We assume that $f(p_T)$ is constant\footnote{We take the value of $f(p_T)$ at lowest available $p_T$ for the unavailable $p_T$ values below the thresholds.} below these thresholds.\footnote{Results for identified particles, shown in Fig.~\ref{fig:ALICE} (d)-(f), confirm that after averaging over particle species, the variation of $f(p_T)$ is modest at low $p_T$.}
At high $p_T$, $f(p_T)$ gradually decreases, which is the non-hydrodynamic effect we aim at capturing. 
Note that the decrease of $v_2(p_T)$ with $p_T$ is less strong for the more peripheral centrality window, resulting in a $f(p_T)$ which is similar for both centralities at high $p_T$. 
We do not understand the origin of this effect. 

$f(p_T)$ of identified hadrons is  displayed in Figs.~\ref{fig:ALICE} (d)-(f). 
The estimates using $v_2\{2\}$ and $v_2\{4\}$ are again similar, as in the case of charged hadrons. 
One sees that $f(p_T)$ slightly exceeds unity for pions at very low momentum, $p_T<0.5$~GeV$/c$, in the most central window. 
This effect is likely related to the ``soft pion excess'', i.e., to the fact that hydrodynamics underpredicts pion production at low $p_T$~\cite{Guillen:2020nul}, a phenomenon that could reveal the underlying  dynamics of chiral symmetry~\cite{Grossi:2021gqi,Florio:2021jlx,Bruschke:2025wny}. 
For $p_T>3$~GeV$/c$, $f(p_T)$ is larger for protons than for pions and kaons, which is the consequence of meson-baryon splitting.

\section{Results for $v_0(p_T)$ and $v_{02}(p_T)$}
\label{s:predictions}

We now validate our {\it ad hoc\/} correction procedure using $v_0(p_T)$ data~\cite{ATLAS:2025ztg,ALICE:2025iud}. 
We evaluate $v_0(p_T)$ in the hydrodynamic simulation following closely the experimental procedure.  
For each hydrodynamic event, we first calculate the normalized spectrum $n(p_T)$~\cite{ATLAS:2025ztg} and the transverse momentum per particle $[p_T]$. 
For the latter quantity, we only include particles in the range $0.5<p_T<2$~GeV$/c$ in Fig.~\ref{fig:ATLAS} (c), corresponding to the ATLAS data shown, and particles in the range $0.2<p_T<3$~GeV$/c$ in Fig.~\ref{fig:ALICE} (g)-(i), corresponding to ALICE data shown. 
We then define $v_0(p_T)$ by~\cite{Schenke:2020uqq}
\begin{equation}
  \label{defv0pt}
  v_{0}(p_T) \equiv \frac{\langle n(p_T)[p_T]\rangle-\langle n(p_T)\rangle\langle p_T\rangle}{\langle n(p_T)\rangle\sigma_{p_T}},
\end{equation}
where angular brackets denote the average over events in a centrality class.
$\langle p_T\rangle$ and $\sigma_{p_T}$ are the mean and standard deviation of $[p_T]$ over the class of events. 
We then multiply our result by $f(p_T)$ defined by Eq.~(\ref{defcorrection}). 
We choose $f(p_T)$ from $v_2\{4\}$ as the default value, because this observable is robust with respect to nonflow effects. 
The difference with the estimate from $v_2\{2\}$ is taken into account in the error bar. 

The resulting prediction is compared with ATLAS data for charged hadrons in Fig.~\ref{fig:ATLAS} (c), and with ALICE data for identified hadrons in Fig.~\ref{fig:ALICE} (g)-(h)
Agreement is reasonable for all $p_T$, and significantly better than from viscous hydrodynamic simulations~\cite{Du:2025dpu} or blast-wave models~\cite{Saha:2025nyu}. 
%At high $p_T$, our calculation is below data for the more peripheral centrality window. 
%This is likely explained by nonflow correlations from back-to-back jets, which are larger for more peripheral collisions and bias $v_0(p_T)$ measurements at high $p_T$. 
%We expect that agreement between theory and data will be improved for $v_{02}(p_T)$, which is less sensitive to nonflow effects. 
%$v_0(p_T)$ results . 
%Agreement with data (which are available only for the first two centrality windows) is not perfect, but better than in pure hydrodynamic calculations~\cite{Du:2025dpu} 
In particular, we reproduce the observed meson-baryon splitting. 

Note that $v_0(p_T)$ is negative at low $p_T$ and then becomes positive at higher $p_T$. 
The reason is that $v_0(p_T)$ is a relative fluctuation of the spectrum, so that its integral, weighted with the particle spectrum, must vanish for all centralities. 
This sum rule holds both in data~\cite{ATLAS:2025ztg} and in hydrodynamics~\cite{Parida:2024ckk}. 
Strictly speaking, our $p_T$-dependent correction factor $f(p_T)$ violates the sum rule, in the sense that after the correction is implemented, the sum rule is no longer strictly enforced. 
This is however a small effect, because $f(p_T)$ varies little at low $p_T$. 
The $p_T$ dependence of $f(p_T)$ becomes large for $p_T\gtrsim 3$~GeV$/c$, but there are so few particles in this range that their contribution to the sum rule is  negligible. 

We conclude that our {\it ad hoc\/} correction procedure, despite its imperfections, works rather well for $v_0(p_T)$. 
Since the physics probed by $v_{02}(p_T)$ is similar to that of $v_0(p_T)$~\cite{Parida:2025eqv}, we expect that the level of success will be comparable. 

We finally evaluate the new observable $v_{02}(p_T)$ along the same lines. 
It is defined by~\cite{Parida:2025eqv}
\begin{equation}
  \label{defv02pt}
  v_{02}(p_T) \equiv \frac{\langle n(p_T)v_2^2\rangle-\langle n(p_T)\rangle\langle v_2^2\rangle}{\langle n(p_T)\rangle\langle v_2^2\rangle},
\end{equation}
where $v_2$ is the modulus of the second Fourier coefficient of the azimuthal distribution of outgoing particles in the event, integrated over the momentum range $0.5<p_T<5$~GeV$/c$ in Fig.~\ref{fig:ATLAS} (d) and $0.2<p_T<3$~GeV$/c$ in Fig.~\ref{fig:ALICE} (j)-(l).  
Note that it is similar to $v_0(p_T)$ defined by Eq.~(\ref{defv0pt}), up to the replacement of $[p_T]$ with $v_2^2$. 
We then multiply our result by $f(p_T)$ defined by Eq.~(\ref{defcorrection}). 

At first sight, $v_{02}(p_T)$ roughly resembles $v_0(p_T)$, but is smaller by a factor $\approx 5$. 
This holds both for charged hadrons (Fig.~\ref{fig:ATLAS} (d)) and identified hadrons (Fig.~\ref{fig:ALICE} (j)-(l)). 
In particular, we predict a clear meson-baryon splitting at high $p_T$. 
There are however significant differences between the two observables, beyond the overall normalization:

\begin{itemize}
\item The $p_T$ dependences of $v_{0}(p_T)$ and $v_{02}(p_T)$  differ below 1~GeV$/c$. The difference is more pronounced for larger centrality percentiles, where the variation of $v_{02}(p_T)$ with $p_T$ is not monotonic:  
It first decreases and then increases. 
This effect is also more pronounced for heavier particles, e.g., protons.  
It is a signature of the variation of the spectrum due to the geometry~\cite{Samanta:2026omo}: 
A more elliptic geometry results in stronger pressure gradients, which increase particle production at high $p_T$, without modifying the total number of particles and the mean $p_T$. 
The corresponding contribution to $v_{02}(p_T)$ has two nodes (positive, negative, positive as $p_T$ increases)~\cite{Parida:2025eqv}.
Note that this modification goes along with an inversion of the usual mass ordering at very low $p_T$, where $v_{02}(p_T)$ is larger for heavier particles. 
We predict that the crossing occurs around $p_T\approx 0.9$~GeV$/c$ between protons and kaons, and around  $p_T\approx 0.4$~GeV$/c$ between kaons and pions. 
A hint of this inversion is seen in event-shape-engineered events~\cite{ALICE:2015lib,Parida:2025eqv}. 
\item 
The centrality dependences also differ. 
In absolute magnitude, $v_0(p_T)$ increases with centrality percentile, while there is no such clear trend for $v_{02}(p_T)$, which has a weak maximum around 30-40\% centrality.   
The increase of $v_0(p_T)$ is well understood: 
Initial-state fluctuations are larger (in relative value) for smaller systems, so that they increase with centrality percentile. 
Therefore, {\it flow\/} fluctuations, as measured by $v_0(p_T)$, also increase, even though this increase is somewhat tamed by the viscous suppression of the hydrodynamic response to initial state fluctuations,  which is stronger in peripheral collisions.
The centrality dependence of $v_{02}(p_T)$, which differs from that of $v_0(p_T)$, is a subtle interplay between several effects. 
It can be understood through the sum rule~\cite{Parida:2025eqv}
\begin{equation}
 \int_{p_T} \frac{p_T}{\langle p_T\rangle}\langle n(p_T)\rangle v_{02}(p_T) =
v_0 \, \frac{\sigma_{v_2^2}}{\langle v_2^2\rangle}\,\rho_2,
\end{equation}
where $\rho_2$~\cite{STAR:2024wgy} is the Pearson correlation coefficient between $[p_T]$ and $v_2^2$ (Bo{\.z}ek's correlator~\cite{Bozek:2016yoj}).  
Thus the magnitude of $v_{02}(p_T)$ results from three different factors: 
$v_0\equiv\sigma_{p_T}/\langle p_T\rangle$ (the integrated $v_0(p_T)$) increases with centrality percentile, $\sigma_{v_2^2}/\langle v_2^2\rangle=\sqrt{1-(v_2\{4\}/v_2\{2\})^4}$ (the relative magnitude of elliptic flow fluctuations) has a weak minimum around 20\% centrality~\cite{Giacalone:2017uqx}, and $\rho_2$  {\it decreases\/} with centrality percentile~\cite{Giacalone:2020dln,Schenke:2020uqq}.
Our hydrodynamic simulations give $\rho_2=0.20$ and $0.094$ with the ATLAS kinematic cuts in 10-20\% and 50-60\% centrality windows, in agreement with the latest experimental data~\cite{ATLAS:2022dov}.\footnote{We obtain 
$\rho_2=0.165$, $0.149$, $0.055$ with the kinematic cuts of ALICE  in 10-20\%, 30-40\% and 50-60\% centrality windows, which also seems consistent with recent data~\cite{ALICE:2026cly}.}
\item
For $p_T>6$~GeV$/c$, our simulations predict that $v_{02}(p_T)$ becomes larger in 50-60\% than in 10-20\% centrality (Fig.~\ref{fig:ATLAS} (d)). 
This inversion stems from the slower decrease of $v_2(p_T)$ at high $p_T$ (Fig.~\ref{fig:ATLAS} (a)) in the more peripheral window. 
We do not have a clear physical interpretation of this behaviour.  
Note also that our error bars are large. 
\end{itemize}

\section{Discussion}
We have implemented a new data-driven way of correcting results of hydrodynamic calculations so as to capture generic trends of flow observables seen at high $p_T$, which are not captured by hydrodynamics alone. 
We have modeled the correction as a suppression factor $f(p_T)$ which depends on transverse momentum, but not on the observable under study. 
We have used ideal hydrodynamics as a baseline, for the sake of simplicity, so that $f(p_T)$ accounts both for the viscous suppression and for the non-hydrodynamic features. 
Now, the viscous suppression does depend on the observable under study. 
When we apply the same correction factor to triangular flow, we slightly overpredict the data as illustrated in Fig.~\ref{fig:v3} (Appendix~\ref{fig:v3}), because we evaluate $f(p_T)$ using elliptic flow, and the viscous suppression is larger for  $v_3$ than for $v_2$~\cite{Teaney:2012ke,Gardim:2022vys}. 
Therefore, our approach could be likely improved by using viscous hydrodynamics as a baseline, at the expense of introducing ambiguities due to the poorly-understood hadronization. 
These sophistications are left for future work. 
%We have used elliptic flow data to evaluate $f(p_T)$. 
The validity of the model has been assessed by comparing its predictions with the $v_0(p_T)$ data, yielding good agreement.
On this basis, we have made quantitative predictions for $v_{02}(p_T)$ in several centrality windows. 

We predict that $v_{02}(p_T)$ reaches a maximum value of $\approx 0.02$ for mesons and $\approx 0.03$ for baryons around $p_T\approx 4-5$~GeV/$c$ in Pb+Pb collisions at the LHC, with a weak centrality dependence. Similar to other collective flow observables, the predicted $v_{02}(p_T)$ exhibits a clear meson--baryon splitting at high $p_T$.

The most remarkable feature of $v_{02}(p_T)$ is that at lower $p_T$, its $p_T$ dependence varies with centrality, in contrast with spectra~\cite{Muncinelli:2024izj} and usual flow observables~\cite{ATLAS:2018ezv} ($v_0(p_T)$, $v_2(p_T)$, $v_3(p_T)$). 
The variation with $p_T$ resembles that of $v_0(p_T)$ for central collisions, but gradually becomes non monotonic as the centrality percentile increases. 
The non-monotonicity should be clearly observed for protons  above 30\% centrality, where we predict a $v_{02}(p_T)$ which is positive at very low $p_T$, then turns negative, and becomes positive again above $2$~GeV$/c$. 
This non-trivial centrality dependence is due to the fact that $v_{02}(p_T)$ is the superposition of thermal and geometric components~\cite{Samanta:2026omo}, whose relative weights depend on centrality. 
Confirmation of these predictions in future analyses will provide yet another non-trivial test of the validity of the hydrodynamic picture of ultrarelativistic heavy-ion collisions. 

\begin{acknowledgments}
We thank Anne Sickles and You Zhou for providing us  $v_2\{4\}$ data from ATLAS and ALICE. 
R. Samanta and T. Parida acknowledge 
support from the Polish National Science Centre grant
2023/51/B/ST2/01625.
\end{acknowledgments}

\appendix
\section{Hydrodynamic setup}
\label{s:hydrodetails}

 We simulate 2500 Pb+Pb collisions at $\sqrt{s_{NN}}=5.02$~TeV in each centrality window (10-20\%, 30-40\%, 50-60\%).
We use the initial entropy as a centrality classifier. 
The initial entropy density is given by the \trento{} model of initial conditions~\cite{Moreland:2014oya}, with parameters $p=0$ (initial entropy density profile proportional to $\sqrt{T_AT_B}$), $k=1.4$ (gamma fluctuations), $w=0.6$~fm (nucleon width, no nucleon substructure). 
The initial entropy density profile is normalized so as to reproduce the final multiplicity measured by ALICE in the 0--5\% centrality window~\cite{ALICE:2015juo}.
The hydrodynamic evolution starts at proper time $\tau_0=0.4$~fm$/c$. 
We solve ideal hydrodynamic equations  using the publicly available MUSIC code~\cite{Schenke:2010nt,Schenke:2010rr,Paquet:2015lta}.
The equation of state is taken from Ref.~\cite{Moreland:2015dvc}, which is constructed by matching lattice QCD calculations at zero baryon chemical potential to a hadron resonance gas description in the low-temperature regime. We perform particlization following the Cooper-Frye prescription~\cite{Cooper:1974mv} on a freezeout hypersurface at $T=145$~MeV (corresponding to an energy density $0.18$~GeV/fm$^3$). 

In each event, we evaluate elliptic flow as the complex Fourier coefficient of the azimuthal distribution: $V_2\equiv \langle e^{2i\varphi}\rangle$, where angular brackets denote an average over outgoing hadrons in the event, within the specific $p_T$-interval ($0.2<p_T<3$~GeV$/c$ when considering ALICE data and $0.5<p_T<5$~GeV$/c$ for ATLAS).  
This average can be carried out at fixed $p_T$ (differential flow $V_2(p_T)$) or without any selection in $p_T$ (integrated flow $V_2$). 
$v_2\{2\}(p_T)$ and $v_2\{4\}(p_T)$ are then defined by~\cite{Borghini:2001vi}
\begin{align}
    v_2\{2\}(p_T)&\equiv \frac{\langle V_2(p_T)V_2^*\rangle }{\langle |V_2|^2\rangle^{1/2}}\\
        v_2\{4\}(p_T)&\equiv \frac{2\langle V_2(p_T)V_2^*\rangle \langle |V_2|^2\rangle-\langle V_2(p_T)V_2^*|V_2|^2\rangle}{\left(2\langle |V_2|^2\rangle^2-\langle |V_2|^4\rangle\right)^{3/4}},\nonumber
\end{align}
where angular brackets now denote an average over events in a centrality class. 

\section{Enhanced hydrodynamic results for $v_3$}
\label{s:v3}
\begin{figure}[!htbp]
    \centering
    \includegraphics[width=\linewidth]{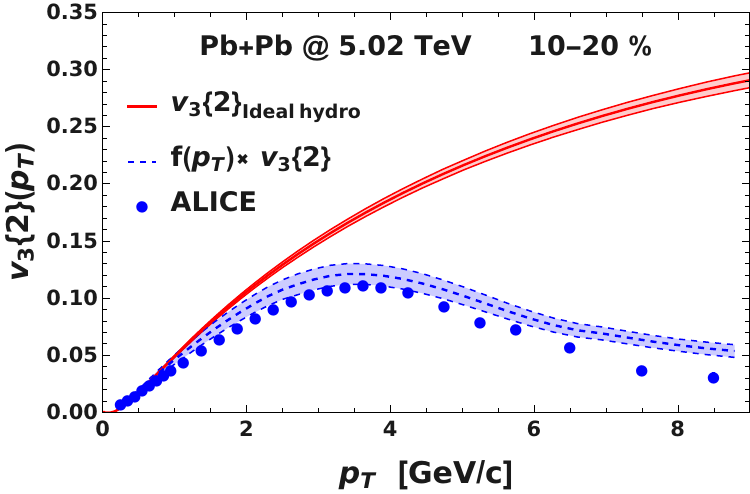}
    \caption{
    Symbols: $v_3\{2\}(p_T)$ for unidentified charged hadrons in 10-20\% centrality~\cite{ALICE:2018rtz}. 
    Lines display the ideal hydrodynamic calculation before and after applying the correction factor $f(p_T)$ defined by Eq.~(\ref{defcorrection}). 
    }
\label{fig:v3}    
\end{figure}
For the sake of illustration, Fig.~\ref{fig:v3} displays the application of our correction procedure to triangular flow data, $v_3\{2\}$. 
We reproduce the $p_T$ dependence of $v_3(p_T)$ in mid-central collisions fairly well  all the way up to 9~GeV$/c$, in the same way as for $v_0(p_T)$ in Fig.~\ref{fig:ATLAS}~(c). 
Ideal hydrodynamics slightly overestimates experiment even after the correction factor $f(p_T)$ is applied. 
At low $p_T$, this is explained by viscous effects, which are the main effect contributing to the correction $f(p_T)$.  
Now, $f(p_T)$ is evaluated using $v_2$, and viscous suppression is larger for $v_3$ than for $v_2$ by a factor $\sim 2$~\cite{Gardim:2022vys}. 
This explains why the corrected results are half-way between ideal hydrodynamics and data for $p_T<2$~GeV$/c$. 
For $p_T>6$~GeV$/c$, the discrepancy between our corrected calculation and data becomes larger. 
This can be ascribed to nonflow effects from back-to-back jets, typically at relative azimuthal angle $\Delta\varphi\approx\pi$, which give a negative contribution to $v_3\{2\}$ since $\cos 3\pi<0$. 
Agreement with data is of similar quality for identified particles (not shown) but worse for larger centrality percentiles where the viscous suppression is larger~\cite{Gardim:2020mmy} (not shown).

\bibliography{v02pt_pred}
\end{document}